%% file: main.tex
\documentclass[twocolumn,aps,prl,reprint,superscriptaddress,floatfix,longbibliography]{revtex4-1}
\pdfoutput=1
\usepackage[latin9]{inputenc}
\usepackage{mathtools}
\usepackage{dsfont}
\usepackage{amsmath}
\usepackage{amsthm}
\usepackage{amssymb}
\usepackage{graphicx}

\makeatletter
\theoremstyle{plain}
\newtheorem{thm}{\protect\theoremname}
\theoremstyle{plain}
\newtheorem{assumption}[thm]{\protect\assumptionname}

\usepackage[dvipsnames]{xcolor}
\usepackage[colorlinks,linkcolor=OrangeRed,urlcolor=NavyBlue,citecolor=RoyalBlue] {hyperref}

\newcommand{\beginsupplement}{%
\pagebreak
\widetext
\setcounter{equation}{0}
\setcounter{figure}{0}
\setcounter{table}{0}
\setcounter{page}{1}
\makeatletter
\renewcommand{\theequation}{S\arabic{equation}}
\renewcommand{\thefigure}{S\arabic{figure}}
}%

\makeatother

\providecommand{\assumptionname}{Assumption}
\providecommand{\theoremname}{Theorem}

\begin{document}
\global\long\def\E{\mathrm{e}}%
\global\long\def\D{\mathrm{d}}%
\global\long\def\I{\mathrm{i}}%
\global\long\def\mat#1{\mathsf{#1}}%
\global\long\def\cf{\textit{cf.}}%
\global\long\def\ie{\textit{i.e.}}%
\global\long\def\eg{\textit{e.g.}}%
\global\long\def\vs{\textit{vs.}}%
 
\global\long\def\ket#1{\left|#1\right\rangle }%

\global\long\def\etal{\textit{et al.}}%
\global\long\def\Tr{\text{Tr}\,}%
 
\global\long\def\im{\text{Im}\,}%
 
\global\long\def\re{\text{Re}\,}%
 
\global\long\def\bra#1{\left\langle #1\right|}%
 
\global\long\def\braket#1#2{\left.\left\langle #1\right|#2\right\rangle }%
 
\global\long\def\obracket#1#2#3{\left\langle #1\right|#2\left|#3\right\rangle }%
 
\global\long\def\proj#1#2{\left.\left.\left|#1\right\rangle \right\langle #2\right|}%
\global\long\def\mds#1{\mathds{#1}}%

\title{Absence of localization in interacting spin chains with a discrete
symmetry}
\author{Benedikt Kloss}
\affiliation{Center for Computational Quantum Physics, Flatiron Institute, 162
Fifth Ave, New York, NY 10010, USA}
\author{Jad C.~Halimeh}
\affiliation{Department of Physics and Arnold Sommerfeld Center for Theoretical
Physics (ASC), Ludwig-Maximilians-Universität München, Theresienstraße
37, D-80333 München, Germany}
\affiliation{Munich Center for Quantum Science and Technology (MCQST), Schellingstraße
4, D-80799 München, Germany}
\author{Achilleas Lazarides}
\affiliation{Interdisciplinary Centre for Mathematical Modelling and Department
of Mathematical Sciences, Loughborough University, Loughborough, Leicestershire
LE11 3TU, UK}
\author{Yevgeny Bar~Lev}
\affiliation{Department of Physics, Ben-Gurion University of the Negev, Beer-Sheva
84105, Israel}
\begin{abstract}
We prove that spin chains symmetric under a combination of mirror
and spin-flip symmetries and with a nondegenerate spectrum show finite
spin transport at zero total magnetization and infinite temperature.
We demonstrate this numerically using two prominent examples: the
Stark many-body localization system (Stark-MBL) and the symmetrized
many-body localization system (symmetrized--MBL). We provide evidence
of delocalization at all energy densities and show that the delocalization
mechanism is robust to breaking the symmetry. We use our results to
construct two localized systems which, when coupled, delocalize each
other.
\end{abstract}
\date{\today}

\maketitle
\emph{Introduction}.---One of the basic assumptions of classical
or quantum statistical mechanics is that interacting many-body systems
thermalize, approaching local thermodynamic equilibrium under unitary
dynamics. This assumption is not satisfied for localized systems,
in which transport is arrested. Two well-known examples are strongly
disordered or ``many-body localized'' (MBL) systems \citep{Basko2006,Gornyi2005,Nandkishore2014,Abanin2018,Alet2018},
and clean systems with a strong tilted potential (``Stark-MBL'')
\citep{vanNieuwenburg2019,Schulz2019}. Significant suppression of
dynamics was experimentally observed in both MBL \citep{Schreiber2015,Choi2016}
and Stark-MBL systems \citep{Scherg2021,morong2021:ObservationStarkManybody}.

Intrinsic instability in localized \emph{noninteracting} systems occurs
due to resonances, which are distinct regions in space with close
energies of the single-particle states \citep{Anderson1958b}. The
probability to have a resonance between two distinct regions decays
exponentially with the distance between them \citep{mott1968:ConductionNoncrystallineSystems,mott1970:ConductionNonCrystallineSystems}.
Lowering the disorder strength increases the density of the resonances
and eventually leads to delocalization at $d\geq3$ \citep{Anderson1958b}.
The resonances also give the dominant contribution to ac conductivity
in localized systems \citep{mott1968:ConductionNoncrystallineSystems,mott1970:ConductionNonCrystallineSystems,Berezinskii1974,Abrikosov1978,Sivan1987}.

For \emph{interacting} systems, the resonant condition includes also
the local interaction energy. Similarly to the noninteracting case
\citep{mott1968:ConductionNoncrystallineSystems,mott1970:ConductionNonCrystallineSystems}
resonances can induce a nonlocal response \citep{khemani2015:NonlocalAdiabaticResponse}.
Nevertheless, under the assumption that the levels of the many-body
spectrum do not attract each other, it was rigorously shown that the
many-body resonances cannot delocalize one-dimensional disordered
systems \citep{Imbrie2014,Imbrie2016}. The proof does not apply for
higher dimensions, and it is currently unclear if localization is
possible for two and higher dimensional interacting systems \citep{roeck2017:ManybodyLocalizationStability,potirniche2019:ExplorationStabilityManybody}.

\begin{figure}
\begin{centering}
\includegraphics{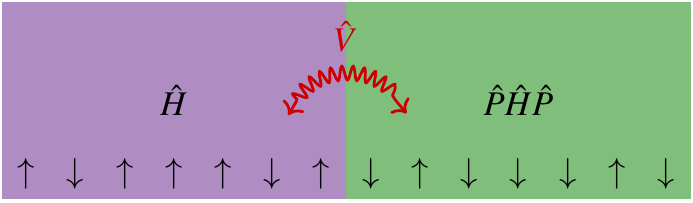}
\par\end{centering}
\caption{Illustration of a symmetrized Hamiltonian. Here $\hat{H}$ conserves
the total magnetization and $\hat{V}$ represents the coupling between
$\hat{H}$ and $\hat{P}\hat{H}\hat{P}$, such that $\hat{P}\hat{V}\hat{P}=\hat{V}.$
The symmetry generator $\hat{P}$ mirrors and flips the spin-pattern
on the left.\label{fig:Illustration}}
\end{figure}
The discussion above refers to disordered systems without global symmetries.
On the other hand, systems with symmetries can have either exact or
a nearly degenerate spectrum, such that the rigorous proof of localization
does not apply \citep{Imbrie2014,Imbrie2016}. Discrete compact symmetries
do not seem to affect localization \citep{huse2013:LocalizationprotectedQuantumOrder,Pekker2014,Kjall2014,vasseur2016:ParticleholeSymmetryManybody,friedman2018:LocalizationprotectedOrderSpin},
however exact resonances can lead to delocalization in translation-invariant
systems \citep{michailidis2018:SlowDynamicsTranslationinvariant,Schulz2019},
as also systems with continuous non-Abelian symmetries \citep{potter2016:SymmetryConstraintsManybody,protopopov2017:EffectSUSymmetry}.
Symmetry-assisted delocalization is however not stable to the addition
of symmetry-breaking perturbations that lift many of the exact resonances
\citep{srivatsa2020:ManyBodyDelocalizationEmergent}.

A number of studies argue that MBL is unstable to the existence of
delocalized inclusions, ruling out the existence of a mobility edge
\citep{deroeck2016:AbsenceManybodyMobility}, and even the MBL transition
itself \citep{deroeck2017:StabilityInstabilityDelocalization,huveneers2017:ClassicalQuantumSystems,thiery2018:ManyBodyDelocalizationQuantum}.
This delocalization mechanism was numerically explored by embedding
of thermal regions in MBL systems \citep{luitz2017:HowSmallQuantum,ponte2017:ThermalInclusionsHow,potirniche2019:ExplorationStabilityManybody},
or by coupling the system to a Markovian bath \citep{sels2022:BathinducedDelocalizationInteracting,morningstar2022:AvalanchesManybodyResonances}.
It is not clear if a similar mechanism is present in clean localized
systems such as Stark-MBL.

In this Letter we prove that localization is absent in a large class
of many-body spin systems with a nondegenerate spectrum. This class
consists of all systems symmetric under the combination of spatial
mirroring and spin flipping. By numerically verifying that the nondegeneracy
assumption is fulfilled for interacting Stark-MBL and appropriately
symmetrized disordered problems, we thus rule out localization in
these systems and then explore the stability of these results to symmetry-breaking
perturbations. Finally, we utilize our result to construct two localized
systems that delocalize each other.

\textit{General argument.--- }\textit{\emph{We consider a spin chain
of length $L$ described by a Hamiltonian $\hat{H}$, and assume the
following:}}
\begin{assumption}
\label{assu:nondeg} $\hat{H}$ has a nondegenerate spectrum.
\end{assumption}

\begin{assumption}
\label{assu:sz-conservation}Total magnetization is conserved, $\left[\hat{H},\sum_{i}\hat{S}_{i}^{z}\right]=0$.
\end{assumption}

\begin{assumption}
\label{assu:P-symmetry}The Hamiltonian is symmetric under a combination
of a mirror symmetry and a spin-flip symmetry defined as
\begin{align}
\hat{P}\hat{S}_{i}^{z}\hat{P} & =-\hat{S}_{L-i+1}^{z},\nonumber \\
\hat{P}\hat{S}_{i}^{\pm}\hat{P} & =\hat{S}_{L-i+1}^{\mp},\label{eq:parity-symmetry}
\end{align}
where $\hat{S}_{i}^{z}$ are spin operators of arbitrary spin size
at site $i$, and $\hat{S}_{i}^{\pm}$ are their corresponding raising
(lowering) operators.
\end{assumption}

Since $\hat{P}^{2}=\hat{\mathds{1}}$ its eigenvalues are $\pm1$.
The commutator $\big[\hat{P},\sum_{i}\hat{S}_{i}^{z}\big]\neq0$ unless
the total magnetization vanishes, and therefore we project the Hamiltonian
onto the zero total magnetization sector, namely, we work in the zero
magnetization sector.

We study spin transport by creating a spin excitation at site $j$
on top of some equilibrium state $\hat{\rho}$, such that $\left[\hat{\rho},\hat{H}\right]=0$,
and assess its spreading using the connected spin-spin correlation
function, 
\begin{equation}
G_{ij}^{\rho}\left(t\right)=\left\langle \hat{S}_{i}^{z}\left(t\right)\hat{S}_{j}^{z}\right\rangle -\left\langle \hat{S}_{i}^{z}\right\rangle \left\langle \hat{S}_{j}^{z}\right\rangle ,\label{eq:spin-spin-connected}
\end{equation}
where $\left\langle \hat{O}\right\rangle \equiv\Tr\left(\hat{\rho}\hat{O}\right)$.
Taking the infinite-time average, $\overline{G_{ij}^{\rho}}=\lim_{T\to\infty}\frac{1}{T}\int_{0}^{T}d\bar{t}G_{ij}^{\rho}\left(\bar{t}\right)$,
and using Assumption~\ref{assu:nondeg} we obtain,
\begin{equation}
\overline{G_{ij}^{\rho}}=\sum_{\alpha}p_{\alpha}\bra{\alpha}\hat{S}_{i}^{z}\ket{\alpha}\bra{\alpha}\hat{S}_{j}^{z}\ket{\alpha}-\left\langle \hat{S}_{i}^{z}\right\rangle \left\langle \hat{S}_{j}^{z}\right\rangle ,\label{eq:spin-spin-connected-inftime}
\end{equation}
where $0\leq p_{\alpha}\leq1$ are the eigenvalues of $\hat{\rho}$.
For systems without spin transport, the spin excitation is expected
to be localized in the vicinity of site $j$ at infinite times, $\overline{G_{ij}^{\rho}}-G_{ij}^{\rho}\left(t=0\right)\sim\exp\left[-\left|i-j\right|/\xi\right]$,
where $\xi$ is the localization length \footnote{Note that, since $\hat{H}$ is symmetric with respect to $\hat{P}$,
we have that $\overline{G_{ij}^{\rho}}=-\overline{G_{\tilde{i}j}^{\rho}}$
where $\tilde{i}=L-i+1$ is the mirrored coordinate. Therefore, localization
of the excitation around $j$ implies also localization around $\tilde{j}$,
which can be arbitrarily distant from $j$. This however does not
mean that the system is delocalized. While the spin excitation can
move for arbitrary distances $\left|j-\tilde{j}\right|$, this is
similar to resonant transfer between site $j$ and site $\tilde{j}$,
which leaves the rest of the system localized. There is no transport
in general. A similar situation occurs in the Anderson insulator \citep{mott1968:ConductionNoncrystallineSystems}
and MBL systems \citep{khemani2015:NonlocalAdiabaticResponse}.}. For systems that relax to equilibrium $\overline{G_{ij}^{\rho}}\to0$
such that the excitation is uniformly spread over the lattice. Here
the process is inherently many-body since it is \emph{not} present
for systems which can be mapped to noninteracting fermions \citep{SM}.
To quantify the spreading of the excitation we use the mean-squared
displacement (MSD), 
\begin{equation}
\sigma_{\rho}^{2}\left(t\right)=\sum_{i=1}^{L}\left(i-j\right)^{2}\left[G_{ij}^{\rho}\left(t\right)-G_{ij}^{\rho}\left(0\right)\right],\label{eq:MSD-rho-dynamics}
\end{equation}
and its corresponding infinite-time average $\overline{\sigma_{\rho}^{2}}$.
For delocalized states, the infinite-time averaged MSD scales as $\overline{\sigma_{\rho}^{2}}\sim L^{2}$,
while for localized states $\overline{\sigma_{\rho}^{2}}\sim\xi^{2}$.
We now prove that for systems satisfying the assumptions above, $\overline{\sigma_{\rho}^{2}}\sim L^{2}$,
implying that at least a \emph{finite} fraction of eigenstates are
delocalized. For brevity, we only provide the sketch of the proof
here; see Supplemental Material (SM) for details \citep{SM}.

We take $\hat{\rho}=\hat{\mathds{1}}/\mathcal{N}$ where $\mathcal{N}=\binom{L}{L/2}$
is the Hilbert space dimension. This corresponds to setting $p_{\alpha}=1/\mathcal{N}$
in Eq.~(\ref{eq:spin-spin-connected-inftime}), such that the infinite-time
average of (\ref{eq:MSD-rho-dynamics}) becomes,
\begin{align}
\overline{\sigma_{\infty}^{2}} & =\frac{1}{\mathcal{N}}\sum_{i=1}^{L}\left(i-j\right)^{2}\sum_{\alpha}\bra{\alpha}\hat{S}_{i}^{z}\ket{\alpha}\bra{\alpha}\hat{S}_{j}^{z}\ket{\alpha}\nonumber \\
 & -\frac{1}{\mathcal{N}}\sum_{i=1}^{L}\left(i-j\right)^{2}\sum_{\alpha}\bra{\alpha}\hat{S}_{i}^{z}\hat{S}_{j}^{z}\ket{\alpha}.\label{eq:MSD-ifntime}
\end{align}
We first note that $\mathcal{N}^{-1}\sum_{\alpha}\bra{\alpha}\hat{S}_{i}^{z}\hat{S}_{j}^{z}\ket{\alpha}=\tfrac{1}{4\left(L-1\right)}$
and therefore the second term in (\ref{eq:MSD-ifntime}) is $O\left(L^{2}\right)$
\citep{SM}. To bound the first term we use the symmetry $\hat{P}$
and the identity,
\begin{equation}
\sum_{i=1}^{L}\left(i-j\right)^{2}\bra{\alpha}\hat{S}_{i}^{z}\ket{\alpha}=\left(\tilde{j}-j\right)\obracket{\alpha}{\hat{D}}{\alpha},\label{eq:MSD_proof_order_reduction}
\end{equation}
where $\hat{D}=\sum_{i}i\hat{S}_{i}^{z}$ is the dipole operator and
$\tilde{i}=L-i+1$ the mirrored coordinate. Inserting this identity
into the first term in (\ref{eq:MSD-ifntime}) and using a combination
of triangle and Hölder inequalities, we bound{\small{}
\begin{align}
\frac{1}{\mathcal{N}}\left|\sum_{i=1}^{L}\left(i-j\right)^{2}\sum_{\alpha}\bra{\alpha}\hat{S}_{i}^{z}\ket{\alpha}\bra{\alpha}\hat{S}_{j}^{z}\ket{\alpha}\right| & \leq\frac{\left|\tilde{j}-j\right|}{2}\left(\frac{1}{\mathcal{N}}\Tr\hat{D}^{2}\right)^{1/2}.
\end{align}
}Since the second term in (\ref{eq:MSD-ifntime}) can be exactly evaluated
and scales as $L^{2}$ and it can be shown that $\left\langle \hat{D}^{2}\right\rangle ^{1/2}=O\left(L^{3/2}\right)$
then for all $\left|j-\tilde{j}\right|<O\left(L^{1/2}\right)$ the
second term is dominating in the thermodynamic limit which yields
$\overline{\sigma_{\infty}^{2}}\sim L^{2}$. It is important to note
that this is \emph{not} an upper bound on $\overline{\sigma_{\infty}^{2}}$
but an asymptotic result, which implies delocalization of a finite
fraction of eigenstates\footnote{Ideally we would like to prove that $\overline{\sigma_{\infty}^{2}}\sim L^{2}$
for all excitation sites $1\leq j\leq L$ and not only sites found
at distance $O\left(L^{1/2}\right)$ from the center of the lattice.
For the models we have considered numerically delocalization holds
for all $j$.}.

The proof does not rule out localization in noninteracting Stark or
Anderson problems, since due to the $\hat{P}$ symmetry there are
degeneracies in the many-body spectrum invalidating Assumption~\ref{assu:nondeg}
\citep{SM}.

In what follows, we numerically demonstrate that Assumption~\ref{assu:nondeg}
is satisfied for two cornerstone models of localization in interacting
systems and provide evidence of delocalization for all energy densities.

\begin{figure}[t!]
\centering{}\includegraphics[width=1\columnwidth]{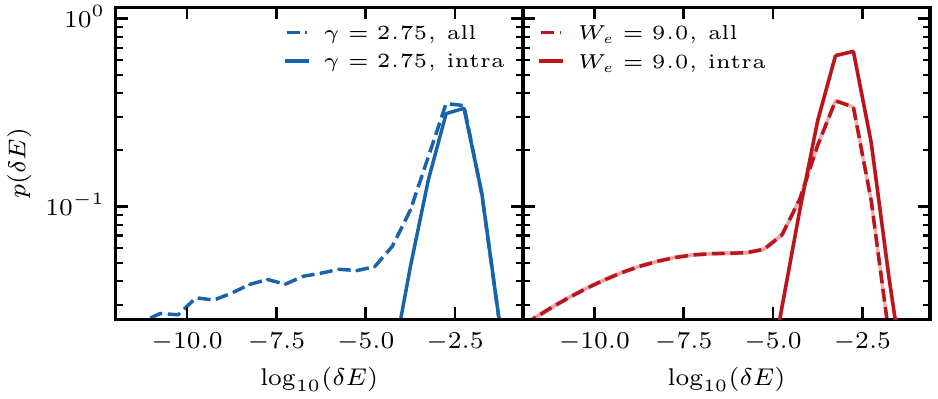}
\caption{\label{fig:spacings}(Color online) Distribution of level spacings
$L=18$ on a log-log scale for the Stark-MBL Hamiltonian, $\gamma=2.75$
(left panel) and the symmetrized--MBL Hamiltonian, $W=9$ (right
panel). Spacings restricted to the even parity symmetry sector (solid
lines) and within the entire zero magnetization sector (dashed lines).
Statistical errors are denoted by line width.}
\end{figure}

\textit{Applications.---}We consider the Hamiltonian of a spin-1/2
chain of length $L$,{\small{}
\begin{align}
\hat{H}= & \sum_{n=1}^{L-1}\bigg[\frac{J}{2}\left(\hat{S}_{n}^{+}\hat{S}_{n+1}^{-}+\hat{S}_{n}^{-}\hat{S}_{n+1}^{+}\right)+\Delta\hat{S}_{n}^{z}\hat{S}_{n+1}^{z}\bigg]+\sum_{n=1}^{L}h_{n}\hat{S}_{n}^{z},\label{eq:Ham}
\end{align}
}where $\hat{S}_{n}^{\pm}$, $\hat{S}_{n}^{z}$ are spin-1/2 operators,
$J$ is the strength of the flip-flop term, $\Delta$ is the strength
of the Ising term and $h_{n}$ is an arbitrary magnetic field. For
$h_{n}=-h_{\tilde{n}}$ the Hamiltonian clearly satisfies Assumptions~\ref{assu:sz-conservation}
and \ref{assu:P-symmetry}. In what follows we numerically verify
that Assumption~\ref{assu:nondeg} is also satisfied for our choices
of $h_{n}$. We consider two cases of ostensibly localized interacting
systems: \textbf{(a)} $h_{n}=\gamma\left(n-\tfrac{L+1}{2}\right)$,
such that all the single-particle states of the fermionic model are
known to be localized for any $\gamma$ and for sufficiently large
$\gamma$ the model is expected to be Stark many-body localized (Stark-MBL)
\citep{vanNieuwenburg2019,Schulz2019} . \textbf{(b)} $h_{n}=-h_{\tilde{n}}$,
but otherwise randomly and uniformly distributed in the interval $\left[-W,W\right]$.
We have verified numerically that all the single-particle states are
strongly localized, and have only rare single-particle resonances,
so that the model might be expected to be many-body localized (MBL)
for sufficiently large $W$, by analogy with the standard MBL case
\citep{Basko2006}. We shall call case (b) symmetrized--MBL, as it
obeys the symmetry embodied in Assumption \ref{assu:P-symmetry}.

\begin{figure}[t!]
\centering{}\includegraphics[width=1\columnwidth]{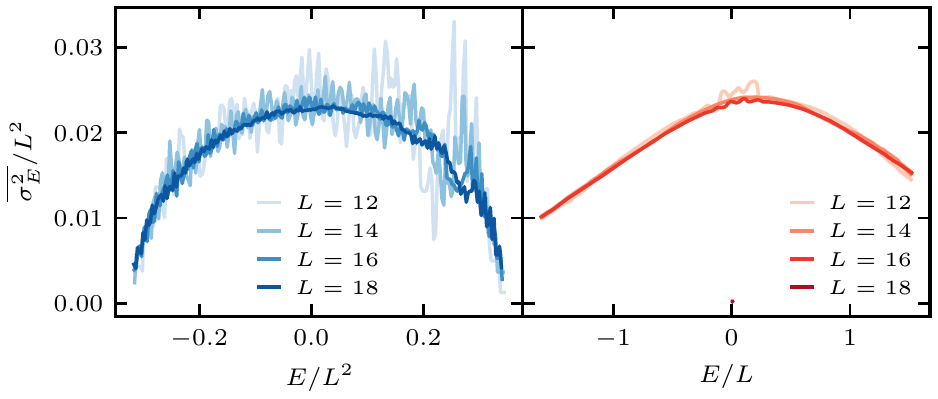}
\caption{\label{fig:sigma_inftime}(Color online) Rescaled infinite-time average
of the microcanonical mean-square displacement $\overline{\sigma_{E}^{2}}/L^{2}$
in the zero-magnetization sector as a function of rescaled energy
for system size $L=12,14,16$, and $18$ (darker shades correspond
to larger systems). The left panel corresponds to the Stark-MBL Hamiltonian
with $\gamma=2.75$ and the energy rescaled as $E/L^{2}$, and the
right panel corresponds to the symmetrized--MBL Hamiltonian with
$W=9$, where the energy is rescaled as $E/L$.}
\end{figure}

We begin by verifying assumption \ref{assu:nondeg} for both models,
by numerically diagonalizing the Hamiltonian for systems sizes $L=11-19$
setting $J=2$ and $\Delta=1$. We work in the zero (1/2) total magnetization
sector for even (odd) system sizes. For the Stark-MBL case we take
$\gamma=2.75$ and for the symmetrized-MBL case $W=9$. We use 10~000
disorder realizations for averaging. Figure~\ref{fig:spacings} shows
a histogram of the logarithm of the eigenvalue spacings, $\log_{10}\delta E$.
Both models have a wide range of eigenvalue pairs that lie very close
to each other compared to the average spacings, but are \emph{no}\textit{t}
degenerate. These quasi-degenerate pairs of states are found across
the symmetry sectors of $\hat{P}$, as we show by restricting the
eigenvalues to the even sector and calculating its distribution (see
also \citep{SM}). The restricted distribution is centered around
the average spacing and does not have a ``fat'' tail stretching
to zero, which characterizes the unrestricted distribution. Since
(\ref{eq:Ham}) satisfies all the assumptions of our proof, we expect
that a finite fraction of its eigenstates are delocalized. To confirm
this, we calculate $\overline{\sigma_{\rho}^{2}}$ in Eq.~(\ref{eq:MSD-ifntime})
within the microcanonical ensemble, $\hat{\rho}\left(E\right)=\mathcal{N}_{E}^{-1}\sum_{\alpha\in I}\ket{\alpha}\bra{\alpha}$,
where $I=\left[E-\Delta E,E+\Delta E\right]$ we take $\Delta E=\frac{\max(E)-\min(E)}{20}$,
and $\mathcal{N}_{E}$ is the number of states in $I$. Figure~\ref{fig:sigma_inftime}
shows that $\overline{\sigma_{E}^{2}}/L^{2}$ plotted vs rescaled
energy for both models is nicely collapsed such that the states at
all energies are delocalized, $\overline{\sigma_{E}^{2}}\sim L^{2}$.

We have established both analytically and numerically that both models
have a delocalized excitation profile at all energy densities and
infinite times. While this result is universal as long as Assumptions~\ref{assu:nondeg}-\ref{assu:P-symmetry}
are satisfied, the temporal and spatial dependence of the excitation
profile (\ref{eq:spin-spin-connected}) are model specific and are
therefore left for the SM \citep{SM}. It is worthwhile to mention
that both models exhibit subdiffusive transport which is better described
by logarithmic subdiffusive transport, $t\sim\ln L$ \citep{Kiefer-Emmanouilidis2020,kiefer-emmanouilidis2021:SlowDelocalizationParticles},
and not power-law subdiffusive transport, $t\sim L^{z}$ ($z>2)$
\citep{Feldmeier2020,zhang2020:SubdiffusionStronglyTilted}. For finite
systems the spin-spin correlation function (\ref{eq:spin-spin-connected})
decays to zero at all sites except $j$ and its mirror $\tilde{j}$,
which suggests a residual memory of initial conditions. The memory,
however, ``fades away'' with increasing system size \citep{SM}.

\begin{figure}[t!]
\centering{}\includegraphics[width=1\columnwidth]{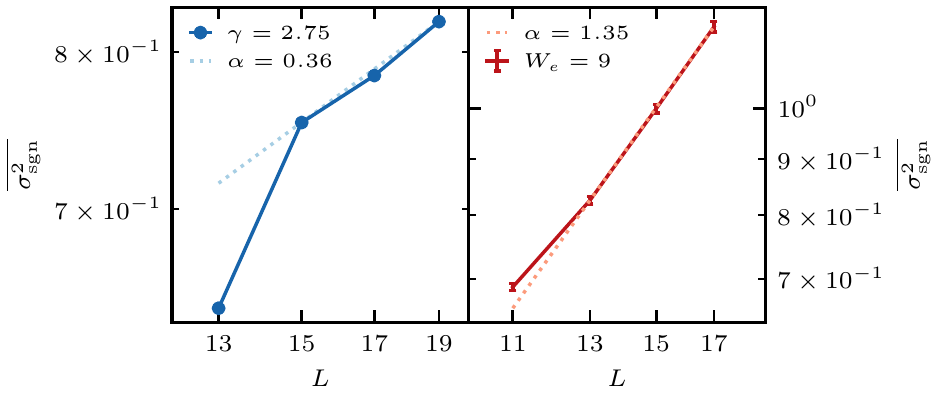}
\caption{(Color online) A log-log plot of $\overline{\sigma_{\text{sgn}}^{2}}$
as a function of $L$ for odd system sizes and total magnetization
1/2. The left panel corresponds to the Stark-MBL Hamiltonian with
$\gamma=2.75$, and the right panel corresponds to the symmetrized--MBL
Hamiltonian with $W=9$.\label{fig:finite_mag}}
\end{figure}

\emph{Symmetry breaking}.---The proof of finite spin transport crucially
depends on existence of the symmetry $\hat{P}$. It is interesting
to see if finite transport persists also when the symmetry is broken.
We have numerically examined a number of ways to break the symmetry
in models described by (\ref{eq:Ham}): taking a finite magnetization,
using an odd system size, or breaking the symmetry of the magnetic
field $h_{n}$. All produce qualitatively similar behavior of dramatically
suppressed dynamics (see for example Refs.~\citep{vanNieuwenburg2019,Schulz2019}).
Here we only present results for odd system sizes and total magnetization
1/2. To examine localization of the excitation profile (\ref{eq:spin-spin-connected}),
we compute a positive version of the MSD by taking $\left|G_{ij}^{\rho}\left(t\right)-G_{ij}^{\rho}\left(0\right)\right|$
in Eq.~(\ref{eq:MSD-rho-dynamics}) and taking an infinite-time average,
$\overline{\sigma_{\text{sgn}}^{2}}$. This is done to avoid the quasi-conservation
of the MSD in Stark-MBL systems \citep{Sala2020,zisling2022:TransportStarkManybody}.
While it implies the absence of diffusion, it does not exclude subdiffusive
transport \citep{Feldmeier2020,zisling2022:TransportStarkManybody}.
Figure~\ref{fig:finite_mag} shows that $\overline{\sigma_{\text{sgn}}^{2}}$
grows with system size for both models. It is hard to extract a reliable
dependence on the system size from the accessible system sizes, but
the growth is consistent with $L^{0.35}$ for the Stark-MBL system
and $L^{1.35}$ for the symmetrized--MBL system. If the growth persists
in the thermodynamic limit it implies asymptotic delocalization. Instead
of breaking the symmetry of the Hamiltonian we can use a nonequilibrium
initial condition $\hat{\rho}$ which either satisfies or breaks the
symmetry. For initial conditions that are odd or even with respect
to $\hat{P}$ (such as the Neél state) we observe some memory of the
initial state, however there is no asymptotic memory retention of
initial conditions that break the symmetry \citep{SM}.

\emph{Discussion}.---We have proved that any spin chain with the
symmetry given in Assumption~\ref{assu:P-symmetry} and a nondegenerate
spectrum exhibits spin transport for a finite measure of its eigenstates.
The proof does not apply to noninteracting systems that have degeneracies
due to the symmetry, and thus can remain localized. We have numerically
demonstrated delocalization of the asymptotic excitation profile for
two cornerstone models of localization in many-body systems: the Stark-MBL
model \footnote{Either purely linear potential or with any additional potential respecting
$h_{i}=-h_{\tilde{i}}$} and the symmetrized--MBL model. Our results suggest that for these
models delocalization happens at all energy densities and spin transport
is subdiffusive, and most probably logarithmic \citep{SM}. Moreover,
our numerical results are consistent with asymptotic delocalization
of the excitation profile also in the case of weak symmetry breaking,
even though the finite-time dynamics is strongly suppressed.

\emph{Constructing a localized delocalizing bath}.---A localized
system is typically a \emph{closed} system with no transport. Coupling
a localized system to a Markovian heat bath induces slow transport
for local coupling \citep{znidaric2016:DiffusiveSubdiffusiveSpin,lezamamergoldlove2022:LogarithmicNoiseinducedDynamics}
or diffusive transport for global coupling \citep{logan1987:DephasingAndersonLocalization,evensky1990:LocalizationDephasingEffects,znidaric2010:DephasinginducedDiffusiveTransport,znidaric2013:TransportDisorderedTightbinding,gopalakrishnan2017:NoiseInducedSubdiffusionStrongly}.
A similar effect is expected to occur if a Markovian bath is replaced
by a sufficiently large thermalizing system. But what if we couple
two localized systems? Is it possible to induce transport in such
a configuration? Since by definition there is transport in neither
system, as a result of the coupling only resonant transfer between
the two systems is possible. A possible guess could be coupling $\hat{H}$
to itself, that we will call the ``$\hat{H}$ to $\hat{H}$'' composite
system. When the two systems are uncoupled all the spectrum is doubly
degenerate and therefore resonant. Under such conditions any small
coupling between the systems lifts the degeneracies and presumably
results in weak transport. However numerical results suggest that
this configuration does \emph{not} result in obvious delocalization
for neither interacting nor noninteracting systems (not shown). \emph{We
conclude that the existence of exact resonances is not a sufficient
condition of delocalization}.

We now use our results to construct a localized system which, when
attached to the edge of a given localized system, described by a localized
Hamiltonian $\hat{H},$ delocalizes it. Since $\hat{H}$ is localized,
the unitarily transformed system $\hat{P}\hat{H}\hat{P}$ is also
localized. For noninteracting systems the symmetry $\hat{P}$ implies
that the single-particle spectrum of $\hat{P}\hat{H}\hat{P}$ is a
reflection around zero of the single-particle spectrum of $\hat{H},$
such that there are no exact \emph{single-particle} resonances \citep{SM}.
On the other hand, the many-body spectrum of $\hat{H}$ is identical
to $\hat{P}\hat{H}\hat{P}$ and therefore has exact many-body resonances,
similarly to the $\hat{H}$ to $\hat{H}$ system. Nevertheless, by
coupling $\hat{H}$ and $\hat{P}\hat{H}\hat{P}$ at the edge using
a symmetric coupling $\hat{P}\hat{V}\hat{P}=\hat{V}$ results in a
composite Hamiltonian that is symmetric under $\hat{P}$: $\hat{H}'=\hat{H}+\hat{P}\hat{H}\hat{P}+\hat{V}$
(see Fig.~\ref{fig:Illustration}). We will call this coupling ``$\hat{H}$
to $\hat{P}\hat{H}\hat{P}$''. Since $\hat{H}'$ satisfies Assumptions~\ref{assu:nondeg}-\ref{assu:P-symmetry}
\footnote{For a coupling which breaks all degeneracies.} it follows
from the delocalization proof that $\hat{H}'$ is delocalized. The
most dramatic demonstration of symmetry-induced delocalization can
be obtained by coupling two Anderson insulators $\hat{H}$ and $\hat{P}\hat{H}\hat{P}.$
A noninteracting coupling of the form $\hat{V}=\hat{S}_{L/2}^{+}\hat{S}_{L/2+1}^{-}+\hat{S}_{L/2}^{-}\hat{S}_{L/2+1}^{+}$
cannot lift the degeneracies and the system is localized. On the other
hand, modifying $\hat{V}$ to include an interacting term, such as
$\hat{S}_{L/2}^{z}\hat{S}_{L/2+1}^{z}$ lifts the degeneracies and
results in delocalization via the delocalization proof. Thus, the
$\hat{H}$ to $\hat{H}$ coupled system has resonances but appears
to be localized, while the $\hat{H}$ to $\hat{P}\hat{H}\hat{P}$
coupling also has resonances, yet is delocalized. Studying the difference
between these systems may provide insight into the role of resonances
in delocalization.

Other open questions are whether the delocalization mechanism carries
over to other spatial symmetries. What is the fastest possible transport
between two coupled localized systems? Is it always logarithmic? It
would be also interesting to see if our delocalization proof can be
generalized to higher dimensions, the microcanonical ensemble, other
conserved quantities, such as the energy, and unbounded local Hilbert
space dimensions. Moreover, implications on thermalization in systems
respecting the symmetry should be also explored.

\bigskip{}

\begin{acknowledgments}
J.C.H.~acknowledges funding from the European Research Council (ERC)
under the European Union's Horizon 2020 research and innovation program
(Grant Agreement no 948141) -- ERC Starting Grant SimUcQuam, and
by the Deutsche Forschungsgemeinschaft (DFG, German Research Foundation)
under Germany's Excellence Strategy -- EXC-2111 -- 390814868. AL
acknowledges support from EPSRC Grant No. EP/V012177/1. This research
was supported by a grant from the United States-Israel Binational
Foundation (BSF, Grant No. 2019644), Jerusalem, Israel, and by the
Israel Science Foundation (grants No. 527/19 and 218/19).
\end{acknowledgments}

\bibliography{lib}

\beginsupplement\include{SM}

\end{document}

%% file: SM.tex
\begin{center}
\global\long\def\E{\mathrm{e}}%
\global\long\def\D{\mathrm{d}}%
\global\long\def\I{\mathrm{i}}%
\global\long\def\mat#1{\mathsf{#1}}%
\global\long\def\cf{\textit{cf.}}%
\global\long\def\ie{\textit{i.e.}}%
\global\long\def\eg{\textit{e.g.}}%
\global\long\def\vs{\textit{vs.}}%
 
\global\long\def\ket#1{\left|#1\right\rangle }%
\par\end{center}

\global\long\def\etal{\textit{et al.}}%
\global\long\def\Tr{\text{Tr}\,}%
 
\global\long\def\im{\text{Im}\,}%
 
\global\long\def\re{\text{Re}\,}%
 
\global\long\def\bra#1{\left\langle #1\right|}%
 
\global\long\def\braket#1#2{\left.\left\langle #1\right|#2\right\rangle }%
 
\global\long\def\obracket#1#2#3{\left\langle #1\right|#2\left|#3\right\rangle }%
 
\global\long\def\proj#1#2{\left.\left.\left|#1\right\rangle \right\langle #2\right|}%
\global\long\def\mds#1{\mathds{#1}}%

\begin{center}
\textbf{\large{}Supplementary Material: Absence of localization in
interacting spin chains with a discrete symmetry}{\large\par}
\par\end{center}

\begin{center}
Benedikt Kloss,$^{1}$ Jad C.~Halimeh,$^{2,3}$ Achilleas Lazarides,$^{4}$
and Yevgeny Bar Lev$^{5}$
\par\end{center}

\begin{center}
$^{1}$\textit{\small{}Center for Computational Quantum Physics, Flatiron
Institute, 162 Fifth Ave, New York, NY 10010, USA}\textit{}\\
$^{2}$\textit{\small{}Department of Physics and Arnold Sommerfeld
Center for Theoretical Physics (ASC), Ludwig-Maximilians-Universität
München, Theresienstraße 37, D-80333 München, Germany}\\
$^{3}$\textit{\small{}Munich Center for Quantum Science and Technology
(MCQST), Schellingstraße 4, D-80799 München, Germany}\\
$^{4}$\textit{\small{}Interdisciplinary Centre for Mathematical Modelling
and Department of Mathematical Sciences, Loughborough University,
Loughborough, Leicestershire LE11 3TU, UK}\\
$^{5}$\textit{\small{}Department of Physics, Ben-Gurion University
of the Negev, Beer-Sheva 84105, Israel}{\small\par}
\par\end{center}

\begin{center}
\today
\par\end{center}

\section{Noninteracting systems}

In this section we examine the properties of noninteracting \emph{fermionic}
systems, which conserve the total particle number and are symmetric
with respect to $\hat{P}$ (see main text). Specifically we consider,
\begin{equation}
\hat{H}=\sum_{i<j}f\left(\left|i-j\right|\right)\hat{c}_{i}^{\dagger}\hat{c}_{j}+h.c.+\sum_{i}v_{i}\left(\hat{n}_{i}-\frac{1}{2}\right),
\end{equation}
where $\hat{c}_{i}^{\dagger}$ creates a fermion at site $i$, $\hat{n}_{i}=\hat{c}_{i}^{\dagger}\hat{c}_{i}$
is the fermion density operator, $f\left(\left|i-j\right|\right)$
corresponds to the hopping rate of the fermions and $v_{i}$ is the
external potential. To show the requirements on $f\left(\left|i-j\right|\right)$
and $v_{i}$ for the Hamiltonian to be symmetric under $\hat{P}$
it is convenient to use the Jordan-Wigner transformation,
\begin{equation}
\hat{c}_{i}^{\dagger}=\left(\prod_{j=1}^{i-1}\hat{\sigma}_{j}^{z}\right)\hat{\sigma}_{i}^{+}\qquad\hat{c}_{i}=\left(\prod_{j=1}^{i-1}\hat{\sigma}_{j}^{z}\right)\hat{\sigma}_{i}^{-}.
\end{equation}
The Hamiltonian written in terms of spins is,
\begin{equation}
\hat{H}=-\sum_{i\neq j}f\left(\left|i-j\right|\right)\sigma_{i}^{+}\left(\prod_{k=i+1}^{j-1}\sigma_{k}^{z}\right)\sigma_{j}^{-}+\frac{1}{2}\sum_{i}v_{i}\sigma_{k}^{z}.
\end{equation}
We see that for,
\begin{align}
f\left(\left|i-j\right|\right) & =0\qquad\left|i-j\right|\,\mod2=1\nonumber \\
v_{i} & =-v_{\tilde{i}},\label{eq:single-particle-props}
\end{align}
the Hamiltonian is symmetric with respect to $\hat{P}$. Using (\ref{eq:single-particle-props})
it is easy to check that the corresponding single-particle Hamiltonian,
\begin{equation}
h_{ij}=f\left(\left|i-j\right|\right)\left(1-\delta_{ij}\right)+v_{i}\delta_{ij},
\end{equation}
is anti-symmetric with respect to the unitary transformation, $\ket i\to\left(-1\right)^{i}\ket{\tilde{i}}$,
which means that the single-particle spectrum is symmetric around
zero. It is important to note that since the single-particle spectrum
is symmetric, there are no exact resonances in the single-body problem,
and therefore creating an excitation at site $j$ does \emph{not}
create a resonant excitation at the mirrored site $\tilde{j}$, as
can be verified numerically. Nevertheless, the many-body spectrum
is degenerate, and therefore Assumption 1 is not satisfied and the
proof we present in Section~\ref{sec:Infinite-time-delocalization}
doesn't apply. Specifically, this includes the noninteracting Stark
and the Anderson problem with the anti-symmetric disorder. We have
verified numerically that these problems indeed remain localized.

\section{Quasi-degeneracies for the Stark-MBL}

The existence of the quasi-degeneracies can be analytically motivated
for the Stark-MBL Hamiltonian,
\begin{equation}
\hat{H}=\sum_{n=1}^{L-1}\bigg[\frac{J}{2}\left(\hat{S}_{n}^{+}\hat{S}_{n+1}^{-}+\hat{S}_{n}^{-}\hat{S}_{n+1}^{+}\right)+\Delta\hat{S}_{n}^{z}\hat{S}_{n+1}^{z}\bigg]+\sum_{n=1}^{L}\gamma n\hat{S}_{n}^{z}.
\end{equation}
We define,
\begin{equation}
\hat{H}_{0}=\Delta\sum_{n=1}^{L-1}\hat{S}_{n}^{z}\hat{S}_{n+1}^{z}+\sum_{n=1}^{L}\gamma n\hat{S}_{n}^{z},
\end{equation}
which is diagonal at the computational basis, namely the eigenbasis
of $\hat{S}_{n}^{z}$ operators. We will treat the flip-flop term
as a perturbation,
\begin{equation}
\hat{V}=\frac{J}{2}\sum_{n=1}^{L-1}\left(\hat{S}_{n}^{+}\hat{S}_{n+1}^{-}+\hat{S}_{n}^{-}\hat{S}_{n+1}^{+}\right).
\end{equation}
Our goal is to show that at zero magnetization there are exponentially
many states which are almost generate. We call a unit-dipole a configuration
which looks like, $d_{i,i+1}^{+}=\left(\uparrow,\downarrow\right)$
or equivalently $d_{i,i+1}^{-}=\left(\downarrow,\uparrow\right)$.
We construct a state $\ket{\psi_{1}}$ by adding to the lattice $N_{+}$
dipoles $d^{+}$ and $N_{-}=L/2-N_{+}$ dipoles . The total dipole
moment of the state is proportional to, $N_{+}-N_{-}$, and its unperturbed
energy is $E_{1}^{\left(0\right)}$. The number of such terms is $\binom{L/2}{N_{+}}$,
namely it is exponential in the size of the system. These states have
typically a different energy due to the interaction $\Delta$. By
applying the symmetry generator $\hat{P}$ we can obtain a new state
$\ket{\psi_{2}}=\hat{P}\ket{\psi_{1}}$, which has the same unperturbed
energy. The operator $\hat{P}$ is a global, while the local perturbation
$\hat{V}$ can only flip one unit-dipole at a time. Therefore an order
of $\alpha L$ such flips are needed to have a non-vanishing coupling
between $\ket{\psi_{2}}$ and $\ket{\psi_{1}},$
\begin{equation}
\obracket{\psi_{1}}{\hat{V}^{\alpha L}\hat{P}}{\psi_{1}}\neq0.
\end{equation}
This means that the degeneracy between the eigenvalues is removed
only at order $\alpha L$ of the perturbation theory, where $\alpha$
is some constant, which depends on the structure of the state. The
resulting splitting between the eigenvalues will be,
\begin{equation}
\delta E_{1}\propto J^{\alpha L}=e^{L\ln J}.
\end{equation}
Since the typical many-body energy spacing is $\delta=\exp\left[-L\ln2\right]$,
the states will appear quasi-degenerate for,
\begin{equation}
\delta E_{1}\ll\delta\qquad J\ll\frac{1}{2}.
\end{equation}

\section{\label{sec:Infinite-time-delocalization}Infinite time delocalization}

We start by calculation of the correlation function $\left\langle \hat{S}_{i}^{z}\hat{S}_{j}^{z}\right\rangle =\mathcal{N}^{-1}\Tr\left(\hat{S}_{i}^{z}\hat{S}_{j}^{z}\right)$
for arbitrary spin size and zero total magnetization. Summing over
all $i$ we have the sum rule
\begin{equation}
\sum_{i}\left\langle \hat{S}_{i}^{z}\hat{S}_{j}^{z}\right\rangle =0.
\end{equation}
Separating the sum to $i=j$ and $i\neq j$ and using the fact that
the expectation value cannot depend on either $i$ or $j$ we get,
\begin{equation}
\sum_{i}\left\langle \hat{S}_{i}^{z}\hat{S}_{j}^{z}\right\rangle =\left\langle \left(\hat{S}_{i}^{z}\right)^{2}\right\rangle +\left(L-1\right)\left\langle \hat{S}_{i}^{z}\hat{S}_{j}^{z}\right\rangle =0.\label{eq:spin-sum-rule}
\end{equation}
Since the total magnetization is set to zero, the total spin $\sum_{i}\vec{S}_{i}$
is rotationally invariant, such that 
\begin{equation}
\left\langle \left(\hat{S}_{i}^{x}\right)^{2}\right\rangle =\left\langle \left(\hat{S}_{i}^{y}\right)^{2}\right\rangle =\left\langle \left(\hat{S}_{i}^{z}\right)^{2}\right\rangle .
\end{equation}
Using this and the definition of the magnitude squared operator of
the spin
\begin{equation}
\left\langle \left(\hat{S}_{i}^{x}\right)^{2}+\left(\hat{S}_{i}^{y}\right)^{2}+\left(\hat{S}_{i}^{z}\right)^{2}\right\rangle =\left\langle \left(\hat{S}_{i}^{2}\right)\right\rangle =s\left(s+1\right),
\end{equation}
where $s$ is the size of the spins, obtain
\begin{equation}
\left\langle \left(\hat{S}_{i}^{z}\right)^{2}\right\rangle =\frac{s\left(s+1\right)}{3}.
\end{equation}
Combining this with the sum-rule (\ref{eq:spin-sum-rule}) we obtain,
\begin{equation}
\left\langle \hat{S}_{i}^{z}\hat{S}_{j}^{z}\right\rangle =\frac{s\left(s+1\right)}{3}\begin{cases}
-\frac{1}{L-1} & i\neq j\\
1 & i=j
\end{cases}.\label{eq:spin-spin-correlator}
\end{equation}

The infinite time average of the MSD, which only assumes that there
are no exact degeneracies is given by,
\begin{equation}
\overline{\sigma_{\infty}^{2}}=\sum_{i=1}^{L}\left(i-j\right)^{2}\frac{1}{\mathcal{N}}\sum_{\alpha}\bra{\alpha}\hat{S}_{i}^{z}\ket{\alpha}\bra{\alpha}\hat{S}_{j}^{z}\ket{\alpha}-\sum_{i=1}^{L}\left(i-j\right)^{2}\left\langle \hat{S}_{i}^{z}\hat{S}_{j}^{z}\right\rangle .\label{eq:inf-msd}
\end{equation}
Using (\ref{eq:spin-spin-correlator}) we see that the last term contributes
\begin{equation}
\sum_{i}\left(i-j\right)^{2}\left\langle \hat{S}_{i}^{z}\hat{S}_{j}^{z}\right\rangle =-\frac{s\left(s+1\right)}{3\left(L-1\right)}\sum_{i}\left(i-j\right)^{2}=-\frac{s\left(s+1\right)}{18}\frac{L\left(\left(L+1\right)\left(2L+1\right)+6j^{2}-6j\left(L+1\right)\right)}{L-1}\sim-\frac{s\left(s+1\right)}{9}L^{2}.
\end{equation}
We now move to the first term which can be simplified,
\begin{align}
\frac{1}{\mathcal{N}}\sum_{\alpha}\sum_{i=1}^{L}\left(i-j\right)^{2}\bra{\alpha}\hat{S}_{i}^{z}\ket{\alpha}\bra{\alpha}\hat{S}_{j}^{z}\ket{\alpha} & =\frac{1}{\mathcal{N}}\sum_{\alpha}\sum_{i=1}^{L}\left(i^{2}+j^{2}-2ij\right)\bra{\alpha}\hat{S}_{i}^{z}\ket{\alpha}\bra{\alpha}\hat{S}_{j}^{z}\ket{\alpha}\nonumber \\
 & =\frac{1}{\mathcal{N}}\sum_{\alpha}\sum_{i=1}^{L}\left(i^{2}-2ij\right)\bra{\alpha}\hat{S}_{i}^{z}\ket{\alpha}\bra{\alpha}\hat{S}_{j}^{z}\ket{\alpha},\label{eq:first-term}
\end{align}
where the last equality follows since we are working a zero magnetization
sector. Focusing on, $\sum_{i=1}^{L}i^{2}\bra{\alpha}\hat{S}_{i}^{z}\ket{\alpha}\bra{\alpha}\hat{S}_{j}^{z}\ket{\alpha}$,
and using the parity symmetry
\begin{align}
\sum_{i=1}^{L}i^{2}\bra{\alpha}\hat{S}_{i}^{z}\ket{\alpha}\bra{\alpha}\hat{S}_{j}^{z}\ket{\alpha} & =\sum_{i=1}^{L}i^{2}\bra{\alpha}\hat{P}\hat{P}\hat{S}_{i}^{z}\hat{P}\hat{P}\ket{\alpha}\bra{\alpha}\hat{S}_{j}^{z}\ket{\alpha}\nonumber \\
 & =-\sum_{i=1}^{L}i^{2}\bra{\alpha}\hat{S}_{L-i+1}^{z}\ket{\alpha}\bra{\alpha}\hat{S}_{j}^{z}\ket{\alpha}.
\end{align}
Changing the summation variables, $i'=L-i+1$ gives,
\begin{align}
\sum_{i=1}^{L}i^{2}\bra{\alpha}\hat{S}_{i}^{z}\ket{\alpha}\bra{\alpha}\hat{S}_{j}^{z}\ket{\alpha} & =-\sum_{i'=1}^{L}\left(L-i'+1\right)^{2}\bra{\alpha}\hat{S}_{i'}^{z}\ket{\alpha}\bra{\alpha}\hat{S}_{j}^{z}\ket{\alpha}\nonumber \\
 & =-\sum_{i=1}^{L}i^{2}\bra{\alpha}\hat{S}_{i}^{z}\ket{\alpha}\bra{\alpha}\hat{S}_{j}^{z}\ket{\alpha}-\left(L+1\right)^{2}\sum_{i=1}^{L}\bra{\alpha}\hat{S}_{i}^{z}\ket{\alpha}\bra{\alpha}\hat{S}_{j}^{z}\ket{\alpha}\nonumber \\
+ & 2\left(L+1\right)\bra{\alpha}\overbrace{\sum_{i}i\hat{S}_{i}^{z}}^{\hat{D}}\ket{\alpha}\bra{\alpha}\hat{S}_{j}^{z}\ket{\alpha},
\end{align}
where the second term vanishes at zero magnetization. We can now rearrange
the terms to obtain the identity,
\begin{equation}
\sum_{i=1}^{L}i^{2}\bra{\alpha}\hat{S}_{i}^{z}\ket{\alpha}\bra{\alpha}\hat{S}_{j}^{z}\ket{\alpha}=\left(L+1\right)\bra{\alpha}\hat{D}\ket{\alpha}\bra{\alpha}\hat{S}_{j}^{z}\ket{\alpha}.
\end{equation}
Inserting this identity into (\ref{eq:first-term}) gives,
\begin{equation}
\frac{1}{\mathcal{N}}\sum_{\alpha}\sum_{i=1}^{L}\left(i-j\right)^{2}\bra{\alpha}\hat{S}_{i}^{z}\ket{\alpha}\bra{\alpha}\hat{S}_{j}^{z}\ket{\alpha}=\left(L+1-2j\right)\frac{1}{\mathcal{N}}\sum_{\alpha}\bra{\alpha}\hat{D}\ket{\alpha}\bra{\alpha}\hat{S}_{j}^{z}\ket{\alpha}=\left(\tilde{j}-j\right)\frac{1}{\mathcal{N}}\sum_{\alpha}\bra{\alpha}\hat{D}\ket{\alpha}\bra{\alpha}\hat{S}_{j}^{z}\ket{\alpha},
\end{equation}
where we have defined the reflected $\tilde{j}\equiv L-j+1$. We note
that for zero magnetization $L$ is even and therefore there is no
$j$ such that $j=\tilde{j}.$ We will now proceed by bounding this
term. Using the triangle inequality,
\begin{equation}
\frac{1}{\mathcal{N}}\left|\sum_{\alpha}\bra{\alpha}\hat{D}\ket{\alpha}\bra{\alpha}\hat{S}_{j}^{z}\ket{\alpha}\right|\leq\frac{1}{\mathcal{N}}\sum_{\alpha}\left|\bra{\alpha}\hat{D}\ket{\alpha}\right|\left|\bra{\alpha}\hat{S}_{j}^{z}\ket{\alpha}\right|,
\end{equation}
now since,
\begin{equation}
\left|\bra{\alpha}\hat{S}_{j}^{z}\ket{\alpha}\right|\leq s,
\end{equation}
we can bound,
\begin{equation}
\frac{1}{\mathcal{N}}\left|\sum_{\alpha}\bra{\alpha}\hat{D}\ket{\alpha}\bra{\alpha}\hat{S}_{j}^{z}\ket{\alpha}\right|\leq\frac{1}{\mathcal{N}}\sum_{\alpha}\left|\bra{\alpha}\hat{D}\ket{\alpha}\right|\left|\bra{\alpha}\hat{S}_{j}^{z}\ket{\alpha}\right|\leq\frac{s}{\mathcal{N}}\sum_{\alpha}\left|\bra{\alpha}\hat{D}\ket{\alpha}\right|.
\end{equation}
The right-hand side is not a trace of a matrix, and depends on the
basis. However for any diagonalizabe matrix we can change to a basis
$\ket n$, where $\hat{D}$ is diagonal, such that,
\begin{equation}
\frac{1}{\mathcal{N}}\sum_{\alpha}\left|\bra{\alpha}\hat{D}\ket{\alpha}\right|=\frac{1}{\mathcal{N}}\sum_{\alpha}\left|\sum_{n.m}\braket{\alpha}n\bra n\hat{D}\ket m\braket m{\alpha}\right|=\frac{1}{\mathcal{N}}\sum_{\alpha}\left|\sum_{n}d_{n}\left|\braket{\alpha}n\right|^{2}\right|,
\end{equation}
using the triangle inequality again we obtain,
\begin{equation}
\frac{1}{\mathcal{N}}\sum_{\alpha}\left|\bra{\alpha}\hat{D}\ket{\alpha}\right|=\frac{1}{\mathcal{N}}\sum_{\alpha}\left|\sum_{n}d_{n}\left|\braket{\alpha}n\right|^{2}\right|\leq\frac{1}{\mathcal{N}}\sum_{\alpha}\sum_{n}\left|d_{n}\right|\left|\braket{\alpha}n\right|^{2}=\frac{1}{\mathcal{N}}\sum_{n}\left|d_{n}\right|,
\end{equation}
which means that $\frac{1}{\mathcal{N}}\sum_{\alpha}\left|\bra{\alpha}\hat{D}\ket{\alpha}\right|$
is maximized in the basis where $\hat{D}$ is diagonal. We now use
Jensen's inequality and obtain finally,
\begin{equation}
\frac{1}{\mathcal{N}}\sum_{\alpha}\left|\bra{\alpha}\hat{D}\ket{\alpha}\right|\leq\frac{1}{\mathcal{N}}\sum_{n}\left|d_{n}\right|\leq\left(\frac{1}{\mathcal{N}}\sum_{n}d_{n}^{2}\right)^{1/2}=\left(\frac{1}{\mathcal{N}}\sum_{n}\Tr\hat{D}^{2}\right)^{1/2}\equiv\left\langle \hat{D}^{2}\right\rangle ^{1/2}.
\end{equation}
The expectation $\left\langle \hat{D}^{2}\right\rangle $ can be evaluated
exactly using (\ref{eq:spin-spin-correlator}),
\begin{align}
\left\langle \hat{D}^{2}\right\rangle  & =\sum_{i,j}i\,j\,\left\langle \hat{S}_{i}^{z}\hat{S}_{j}^{z}\right\rangle =\frac{s\left(s+1\right)}{3}\left[\left[-\frac{1}{L-1}\right]\left(\sum_{i\neq j}i\,j\right)+\sum_{i}i^{2}\right]\nonumber \\
 & =\frac{s\left(s+1\right)}{3}\left[-\frac{\left(\left(\sum_{i=1}^{L}i\right)^{2}-\sum_{i=1}^{L}i^{2}\right)}{L-1}+\sum_{i=1}^{L}i^{2}\right]\nonumber \\
 & =\frac{s\left(s+1\right)}{3}\left[-\frac{L\left(L+1\right)\left(3L+2\right)}{12}+\frac{L\left(L+1\right)\left(2L+1\right)}{6}\right]\nonumber \\
 & =\frac{s\left(s+1\right)}{36}L^{2}\left(L+1\right).
\end{align}
Combining all the results gives,
\begin{equation}
\frac{1}{\mathcal{N}}\left|\sum_{\alpha}\sum_{i=1}^{L}\left(i-j\right)^{2}\bra{\alpha}\hat{S}_{i}^{z}\ket{\alpha}\bra{\alpha}\hat{S}_{j}^{z}\ket{\alpha}\right|\leq s\left|\tilde{j}-j\right|\sqrt{\frac{s\left(s+1\right)}{36}L^{2}\left(L+1\right)}=O\left(L^{3/2}\right).
\end{equation}
Comparing to the second-term of the MSD in (\ref{eq:inf-msd}) we
obtain that for $\left|\tilde{j}-j\right|\leq AL^{1/2}$, where $A>0$
is some constant,
\begin{equation}
\overline{\sigma_{\infty}^{2}}\sim\frac{s\left(s+1\right)}{9}L^{2},
\end{equation}
which concludes the proof that at least a fraction of eigenstates
in the system are delocalized.

\section{Spatial profile of the spin excitation}

\begin{figure}
\includegraphics[width=0.45\textwidth]{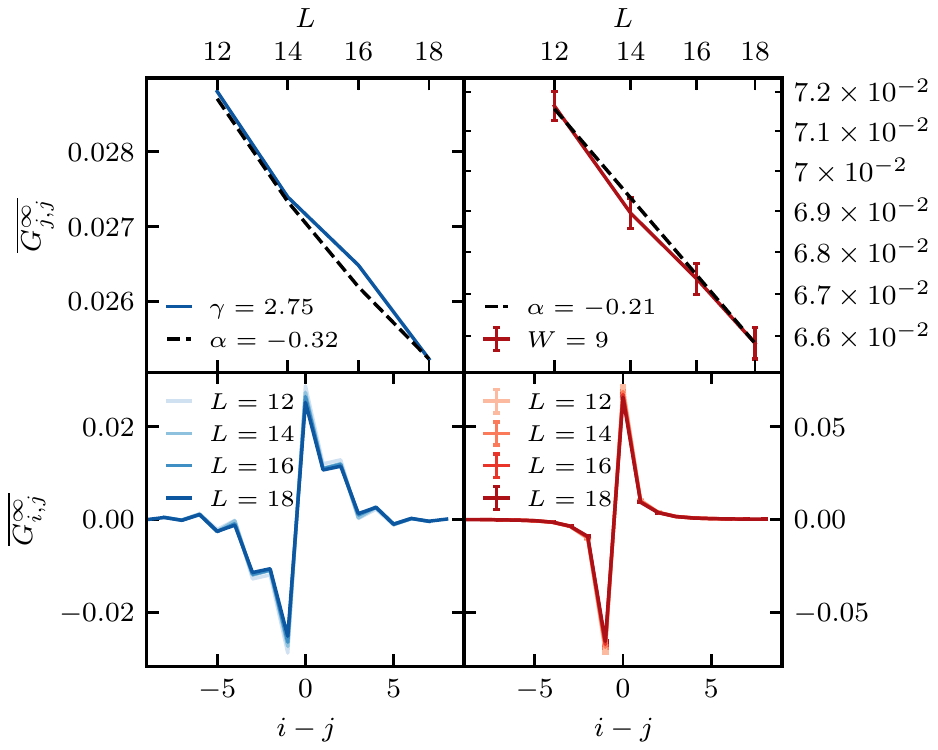}\includegraphics[width=0.45\textwidth]{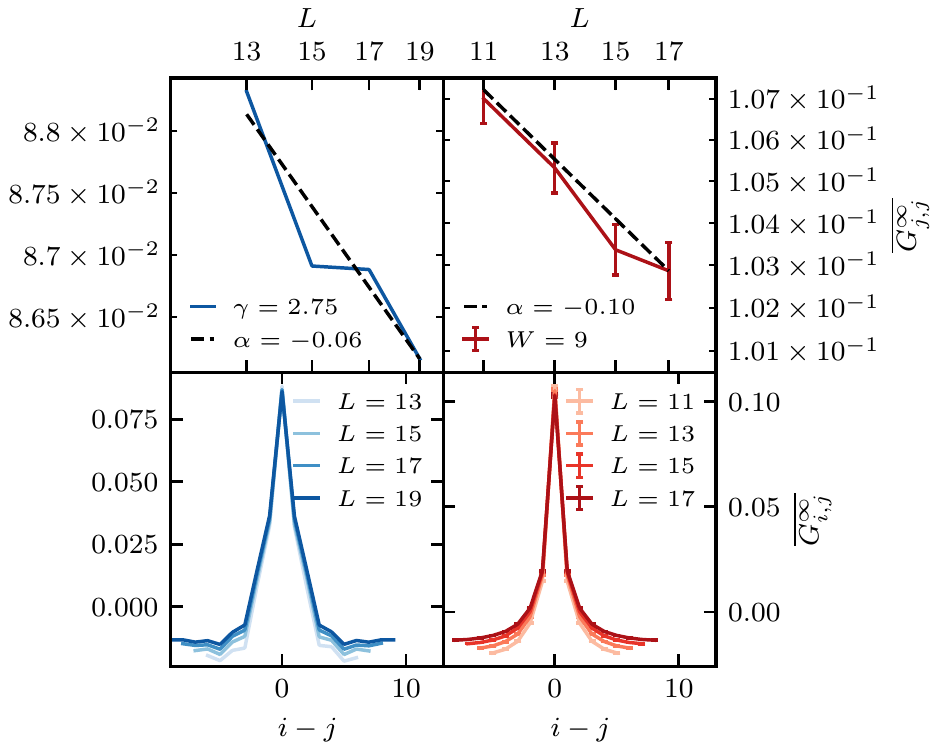}

\caption{\label{fig:autocorr-even-odd}Infinite-temperature, infinite time
average of connected spin-spin correlation function $\overline{G_{i,j}^{\infty}}$,
for Stark-MBL system with $\gamma=2.75$ (left panels) and symmetrized-MBL
system with $W=9$ (right panels) in the zero-magnetization sector.
Top panels show autocorrelator for $j=\lceil L/2\rceil$ as function
of $L$ on log-log scale, while bottom panels show the spatial profile
for several system sizes (darker shades indicate larger systems).}
\end{figure}

The mean-square displacement (MSD) is lacking the spatial information
on the spreading of the spin excitation, which is contained in the
infinite-time averaged spin-spin correlation function,
\begin{equation}
\overline{G_{ij}^{\infty}}=\frac{1}{\mathcal{N}}\sum_{\alpha}\bra{\alpha}\hat{S}_{i}^{z}\ket{\alpha}\bra{\alpha}\hat{S}_{j}^{z}\ket{\alpha}.
\end{equation}
For delocalized systems with no memory of the initial condition this
function is expected to vanish at all sites. In the bottom row of
Fig.~\ref{fig:autocorr-even-odd} we calculate and plot $\overline{G_{ij}^{\infty}}$
for the Stark-MBL and symmetrized--MBL problem for a number of even
(left) and odd (right) system sizes. For even system sizes the total
magnetization is zero and the system is symmetric with respect to
$\hat{P}$, which yields to the anti-symmetric shape $\overline{G_{ij}^{\infty}}=-\overline{G_{\tilde{i}j}^{\infty}}$.
For odd system sizes the total magnetization is 1/2 and the symmetry
$\hat{P}$ is broken, such that the shape of the excitation $\overline{G_{ij}^{\infty}}$
doesn't have to be anti-symmetric. For even system sizes the correlation
function is close to zero at all sites, except $i=j$ and $i=\tilde{j}$,
indicating a relaxation to equilibrium, however for odd system sizes
all sites appear to be away from zero.

The nonzero value of $\overline{G_{ij}^{\infty}}$ for $i=j$ and
$i=\tilde{j}$, indicates some memory of the initial condition, however
as the top row of Fig.~\ref{fig:autocorr-even-odd} shows this memory
is decaying with the system size for both odd and even system sizes.
We cannot reliably extract the dependence of the decay on the system
size, but it is quite slow as one can learn from the qualitative power-law
fits that are listed in the top row of Fig.~\ref{fig:autocorr-even-odd}.
It is important to note that non-vanishing of $\overline{G_{ij}^{\infty}}$
for a finite number of sites, is consistent with finite transport,
since finite transport requires $\overline{\sigma_{\infty}^{2}}\sim L^{2}$
which includes a contribution from an extensive number of sites. Therefore
existence of finite memory is not in contradiction to the proof in
Section.~\ref{sec:Infinite-time-delocalization}. Here the memory
of the excitation appears to fade away in the thermodynamic limit
for both symmetry preserving and symmetry breaking systems.

\section{Excitation spreading}

\begin{figure}
\includegraphics{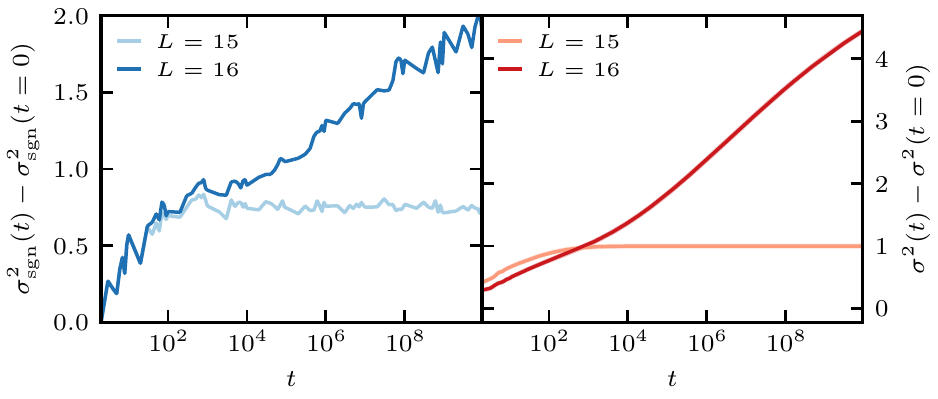}

\caption{\label{fig:Dynamics}Dynamics of the (positive) mean-squared displacement
for Stark-MBL system with $\gamma=2.75$ (left panel) and symmetrized-MBL
system with $W=9$ (right panel), plotted on semilog scale. The darker
shade corresponds to zero-magnetization, while the lighter shade indicates
results from a finite-magnetization sector. The standard deviation
over disorder realizations for the symmetrized MBL is indicated by
shaded areas.}
\end{figure}

In this section we consider the of the time-dependence of the excitation
profile for both Stark-MBL and symmetrized--MBL systems for both
symmetry preserving (even $L)$ and symmetry breaking (odd $L$) system
sizes. For this purpose we compute the positive MSD,
\begin{equation}
\sigma_{\text{sgn}}^{2}\left(t\right)=\sum_{i=1}^{L}\left(i-j\right)^{2}\left|G_{ij}^{\infty}\left(t\right)-G_{ij}^{\infty}\left(0\right)\right|,\label{eq:signed-msd}
\end{equation}
which via triangle inequality bounds $\sigma_{\infty}^{2}\left(t\right)\leq\sigma_{\text{sgn}}^{2}\left(t\right)$.
We take $\sigma_{\text{sgn}}^{2}\left(t\right)$, and not $\sigma_{\infty}^{2}\left(t\right)$,
since for example for dipole preserving systems $\sigma_{\infty}^{2}\left(t\right)\leq C$
uniformly in time while there is still slow subdiffusive transport
\citep{Feldmeier2020,zisling2022:TransportStarkManybody}. 

From Fig.~\ref{fig:Dynamics} we see that while the dynamical behavior
of Stark-MBL and symmetrized--MBL are very similar the long time
behavior of symmetry preserving and symmetry breaking systems is very
different. Symmetry preserving systems have an intermediate plateau
after which the positive MSD grows logarithmically towards its infinite-time
value. The height of the intermediate plateau decreases weakly with
increasing the tilted-field or disorder strengths and delays the approach
to the asymptotic plateau. The occurrence of quasi-degeneracies in
the spectrum of symmetry preserving systems doesn't allow us to numerically
compute the correct dynamics of the system beyond $t>10^{16}$, we
therefore don't present the dynamics beyond these time in Fig.~\ref{fig:Dynamics}.

The dynamics of the positive MSD for symmetry breaking systems follows
the dynamics of symmetry preserving systems up to time $t_{*}$, which
increases with increasing the strength of the tilted field or the
disorder, but does not depend on the system size. Interestingly, the
intermediate plateau of symmetry preserving systems coincides with
the asymptotic plateau of symmetry breaking systems. As explained
in the main text the asymptotic value slowly increases with system
size.

While here we present results of symmetry breaking by going to an
even system size, we have observed very similar phenomenology if the
symmetry is broken differently. For example, by considering a nonzero
magnetization at even $L$ or by adding weak disorder or curvature.

\section{Symmetry breaking of initial conditions}

\begin{figure}
\includegraphics{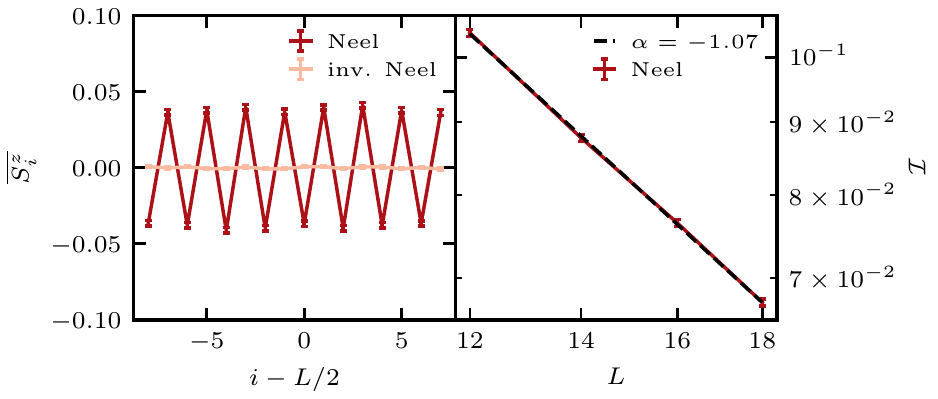}

\caption{\label{fig:Neel}\emph{Left panel}: Infinite-time average of the magnetization
profile starting from initial states related to the Néel state for
the symmetrized-MBL system with $W=9$ and L=16. Results are shown
for the Néel state (dark red), $\protect\ket{\uparrow\downarrow\uparrow\downarrow\uparrow\downarrow\uparrow\downarrow\uparrow\downarrow\uparrow\downarrow\uparrow\downarrow\uparrow\downarrow}$
and inverted Néel state, with the spins on the right half of the lattice
flipped, $\protect\ket{\uparrow\downarrow\uparrow\downarrow\uparrow\downarrow\uparrow\downarrow\downarrow\uparrow\downarrow\uparrow\downarrow\uparrow\downarrow\uparrow}$.
\emph{Right panel}: Asymptotic imbalance of the Néel state as a function
of system size for even system sizes, plotted on a log-log scale.}
\end{figure}

The sensitivity of both Stark-MBL and symmetrized-MBL systems to symmetry
breaking can also be observed via breaking the symmetry in the initial
state and not the Hamiltonian. In the left panel of Fig.~\ref{fig:Neel}
we show the infinite-time average of $\left\langle \hat{S}_{i}^{z}\left(t\right)\right\rangle $,
\begin{equation}
\overline{S_{i}^{z}}=\overline{\left\langle \Psi\left|\hat{S}_{i}^{z}\left(t\right)\right|\Psi\right\rangle },
\end{equation}
for the Néel state $\ket{\text{Néel}}=\ket{\uparrow\downarrow\uparrow\downarrow\uparrow\downarrow\uparrow\downarrow\uparrow\downarrow\uparrow\downarrow\uparrow\downarrow\uparrow\downarrow}$,
which is even under $\hat{P}$. We see that $\overline{S_{i}^{z}}$
shows residual memory of this initial condition. We can quantify this
memory using the imbalance,
\begin{equation}
\mathcal{I}=\sum_{i=1}^{L}\left(-1\right)^{i}\overline{S_{i}^{z}},
\end{equation}
 which slowly decays with the size of the system (see right panel
of Fig.~\ref{fig:Neel}). Flipping the spins in half of the system,
\begin{equation}
\ket{\text{Inverted Néel}}=\prod_{i=L/2+1}^{L}\sigma_{i}^{x}\ket{\text{Néel}}=\ket{\uparrow\downarrow\uparrow\downarrow\uparrow\downarrow\uparrow\downarrow\downarrow\uparrow\downarrow\uparrow\downarrow\uparrow\downarrow\uparrow},
\end{equation}
results in a state which is neither odd nor even under $\hat{P}.$
As can be seen from the left panel of Fig.~\ref{fig:Neel} any memory
of the initial condition for this state is absent already for $L=16$.